\documentclass[aip, rsi, numerical, preprint]{revtex4-1}

\usepackage[pdftex]{graphicx}

\begin{document}

\title{A tunable low-energy photon source for high-resolution angle-resolved photoemission spectroscopy}

\author{John W. Harter}
\affiliation{Laboratory of Atomic and Solid State Physics, Department of Physics, Cornell University, Ithaca, New York 14853, USA}

\author{Philip D. C. King}
\affiliation{Laboratory of Atomic and Solid State Physics, Department of Physics, Cornell University, Ithaca, New York 14853, USA}
\affiliation{Kavli Institute at Cornell for Nanoscale Science, Ithaca, New York 14853, USA}

\author{Eric J. Monkman}
\author{Daniel E. Shai}
\author{Yuefeng Nie}
\author{Masaki Uchida}
\author{Bulat Burganov}
\author{Shouvik Chatterjee}
\affiliation{Laboratory of Atomic and Solid State Physics, Department of Physics, Cornell University, Ithaca, New York 14853, USA}

\author{Kyle M. Shen}
\email[Author to whom correspondence should be addressed: ]{kmshen@cornell.edu}
\affiliation{Laboratory of Atomic and Solid State Physics, Department of Physics, Cornell University, Ithaca, New York 14853, USA}
\affiliation{Kavli Institute at Cornell for Nanoscale Science, Ithaca, New York 14853, USA}

\date{\today}

\begin{abstract}
We describe a tunable low-energy photon source consisting of a laser-driven xenon plasma lamp coupled to a Czerny-Turner monochromator. The combined tunability, brightness, and narrow spectral bandwidth make this light source useful in laboratory-based high-resolution photoemission spectroscopy experiments. The source supplies photons with energies up to $\sim$7 eV, delivering under typical conditions $>10^{12}$ ph/s within a 10 meV spectral bandwidth, which is comparable to helium plasma lamps and many synchrotron beamlines. We first describe the lamp and monochromator system and then characterize its output, with attention to those parameters which are of interest for photoemission experiments. Finally, we present angle-resolved photoemission spectroscopy data using the light source and compare its performance to a conventional helium plasma lamp.
\end{abstract}

\maketitle

\section{Introduction}

Angle-resolved photoemission spectroscopy (ARPES) plays a valuable role in the field of condensed matter physics by offering a direct momentum-space probe of the underlying electronic structure of solids. Motivated by the demand for high-quality data, both the instrumental energy and momentum resolution and the detection efficiency of ARPES systems have improved substantially over the last decades \cite{damascelli2003}. ARPES is based on the photoelectric effect, in which an emitted electron's kinetic energy, $E_{kin}$, is given by
\begin{equation}
\label{photoelectricEffectEqn}
E_{kin} = h\nu - \phi - E_B,
\end{equation}
where $h\nu$ is the incident photon energy, $\phi$ is the work function of the system, and $E_B$ is the binding energy of the electron within the solid \cite{hufner2003}. A key component of photoemission, therefore, is the photon source, which is often a synchrotron because the light must be simultaneously bright and possess a narrow spectral bandwidth in order to maintain a high energy resolution. A common alternative to synchrotrons is the laboratory-based noble gas plasma lamp. A major drawback of plasma lamps is the fact that the photon energy is fixed at a discrete set of atomic lines. For a helium lamp, this encompasses the He-I$\alpha$ line at 21.2 eV and the He-II$\alpha$ line at 40.8 eV.

The spectral lines from plasma lamps, as well as typical photon energies used at synchrotrons, sit close to the minimum of the so-called ``universal curve'' describing the inelastic mean free path of photoexcited electrons within a metal as a function of their kinetic energy \cite{hufner2003,seah1979}. Some benefits of using low-energy light sources therefore include an enhanced sensitivity to more bulk-like properties of the system under study, an increased tolerance to unwanted adsorbates on the sample surface, and a higher momentum resolution. As a result, the development and use of low-energy sources have recently become common in the ARPES community. These are typically laser sources: one such source supplies 7 eV photons by using the nonlinear optical crystal KBe$_2$BO$_3$F$_2$ to generate the second harmonic of a frequency-tripled Nd:YVO$_4$ laser \cite{kiss2008}, and a similar 6 eV light source based on the crystal $\beta$-BaB$_2$O$_4$ generates the fourth harmonic of a titanium-sapphire laser \cite{koralek2007}. Another low-energy source in use employs a low-pressure xenon discharge lamp, which has a number of discrete atomic lines in the energy range 8 -– 11 eV \cite{souma2007,suga2010}.

One major disadvantage of the laser sources described above is a fixed photon energy, which is a particularly important issue at low energies because of the sensitivity to photon wavelength caused by final-state matrix element effects. The ability to tune the photon energy, however, can mitigate this issue. Therefore, it is important to develop a tunable low-energy photon source that can allow for the selection of final states. In this paper we describe a novel laboratory-based low-energy photon source consisting of a laser-driven xenon plasma lamp in conjunction with a Czerny-Turner monochromator. The total cost of the device is less than \$45,000, which is significantly cheaper than noble gas plasma lamps and laser-based systems. Under typical operation, the device delivers $>10^{12}$ ph/s at a 10 meV spectral bandwidth, and its continuous output avoids space-charge complications that are present in pulsed sources. The brightness, energy tunability, and adjustable spectral bandwidth of the light source make it ideally suited for laboratory-based high-resolution ARPES experiments.

\section{The instrument}

\begin{figure}[ht]
\includegraphics{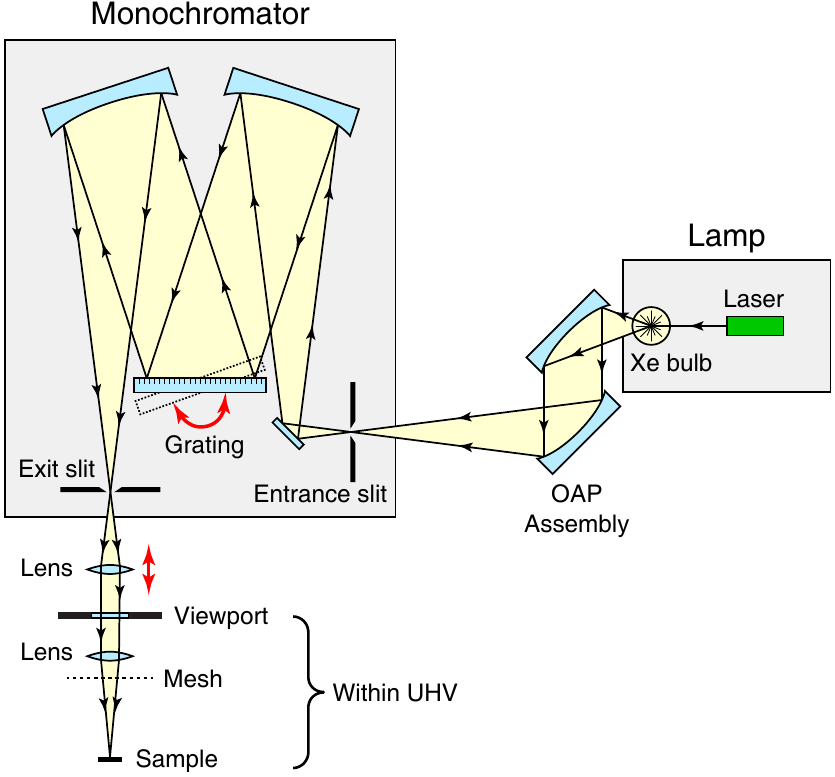}
\caption{\label{figureSchematic}Schematic diagram of the xenon plasma lamp and Czerny-Turner monochromator, as described in the text. Within the lamp housing, an internal laser excites a plasma contained inside the xenon bulb. The light emitted from the plasma is focused onto the entrance slit of the monochromator with a pair of off-axis parabolic mirrors. At the exit slit of the monochromator, a pair of lenses focus the light through an ultra-high vacuum viewport and onto the sample.}
\end{figure}

There are three main components of the photon source: a bright laser-driven xenon plasma lamp, a high-resolution Czerny-Turner monochromator, and a pair of lenses that focus the light onto the sample. As shown in Fig.\ \ref{figureSchematic}, the output of the plasma lamp is directed onto the entrance slit of the monochromator. At the exit slit, the light is guided through an ultra-high vacuum (UHV) viewport into the ARPES vacuum chamber and onto the sample. At the operating energy range of the lamp for photoemission (up to $\sim$7 eV), most optical materials are highly absorptive. Mirrors are therefore ``UV-enhanced'' (MgF$_2$-coated) aluminum; aluminum has a high reflectivity at ultraviolet wavelengths and a MgF$_2$ overlayer prevents oxidation of the metallic surface. Both CaF$_2$ and MgF$_2$ are acceptable materials for lenses and windows, provided they are of ``vacuum grade'' quality, because they have a sufficiently large band gap to allow transmission of photons.

The entire optical beam path of the lamp must be placed within a sealed oxygen-free atmosphere such as dry nitrogen. The ultraviolet light generated by the xenon plasma lamp is capable of generating ozone. This process absorbs a significant fraction of photons, diminishing the brightness of the source. In addition, ozone is harmful to the components of the photon source and will degrade the optical elements within it.

\subsection{The xenon plasma lamp}

The plasma lamp (EQ-1500, Energetiq Technology, Inc.) is a patented commercial instrument featuring an internal continuous wave 980 nm diode laser running at 60 W and driving a plasma within the xenon bulb. The broadband output of the lamp has a divergence of 60$^{\circ}$ and extends from ultraviolet to infrared wavelengths (170 nm to 2100 nm). The electrodeless design of the lamp offers greater stability, a longer lifetime, and higher brightness than xenon or deuterium arc lamps, with a spectral radiance exceeding 10 mW/(mm$^2\cdot$nm$\cdot$sr) over the entire operational wavelength range. The diverging output of the lamp is coupled to an off-axis parabolic (OAP) mirror assembly which focuses the light onto the monochromator entrance slit. The OAP assembly reduces the divergence of the beam to 20$^{\circ}$, with a corresponding magnification of the plasma image by 3$\times$.

\subsection{The Czerny-Turner monochromator}

The monochromator (SP-2355, Princeton Instruments) has a Czerny-Turner configuration with a 300 mm focal length and a UV-optimized holographic plane grating (3600 grooves/mm). The acceptance angle of the instrument is 15$^{\circ}$, receiving roughly half of the total output of the xenon plasma lamp. At the exit slit of the monochromator, the measured linear dispersion of photons is better than 0.33 nm/mm over the entire operating range of the lamp. Based on these parameters, one can compute the required monochromator slit width for a desired spectral bandwidth using the formula
\begin{equation}
\label{slitWidthEqn}
s = \left({hc\over\sigma}\right){\Delta\over{E^2}} \approx \left(3.8\ {\textnormal{mm}\cdot \textnormal{eV}^2\over\textnormal{meV}}\right){\Delta\over{E^2}},
\end{equation}
where $s$ is the slit width (the entrance and exit slits should be matched), $\sigma$ is the linear dispersion of the monochromator, $\Delta$ is the desired spectral bandwidth of the lamp, and $E = hc/\lambda$ is the energy of the photons. For example, at an operating energy of 6 eV, a spectral bandwidth of 5 meV can be achieved with monochromator slits set to 0.5 mm.

\subsection{The focusing lenses}

At the exit slit of the monochromator, a pair of lenses focus the emitted light through a UHV viewport and onto the sample. The focal lengths of these lenses depend on their diameter and on the minimum lens-to-sample distance, which will vary between photoemission systems. As an example, we consider standard lens diameters of 25 mm and a lens-to-sample distance of 130 mm. In order to capture all of the light, the external lens must be within 95 mm of the monochromator exit slit. Choosing focal lengths of 65 mm (external lens) and 130 mm (internal lens), we form an image of the exit slit at the sample position with a 2$\times$ magnification. In the diagram of Fig.\ \ref{figureSchematic}, the focusing lens is shown within the UHV chamber. This is not strictly necessary, but is done in order to minimize the spot size on the sample, as the magnification of the lens system is proportional to the lens-to-sample distance.

An important complication to consider is that of chromatic aberration. The index of refraction of CaF$_2$ has an average dispersion of 9$\times$10$^{-4}$ nm$^{-1}$ in the operating energy range of the photon source \cite{daimon2002}. The focal lengths of the lenses will therefore vary by as much as $\pm$5\% as the photon energy is adjusted. In order to correct for this effect, one must adjust the position of the external lens in order to keep the focus at the sample position. Another complication involves the lens close to the sample, which may charge up from photoemitted electrons. Electrostatic shielding in the form of a grounded mesh between the lens and the sample can prevent the charge field from interfering with the photoemission measurements \cite{read1998}. Alternatively, the focusing lens may be moved outside of the vacuum chamber, at the expense of an increased spot size.

\section{Characterization}

\subsection{Intensity}

\begin{figure}[ht]
\includegraphics{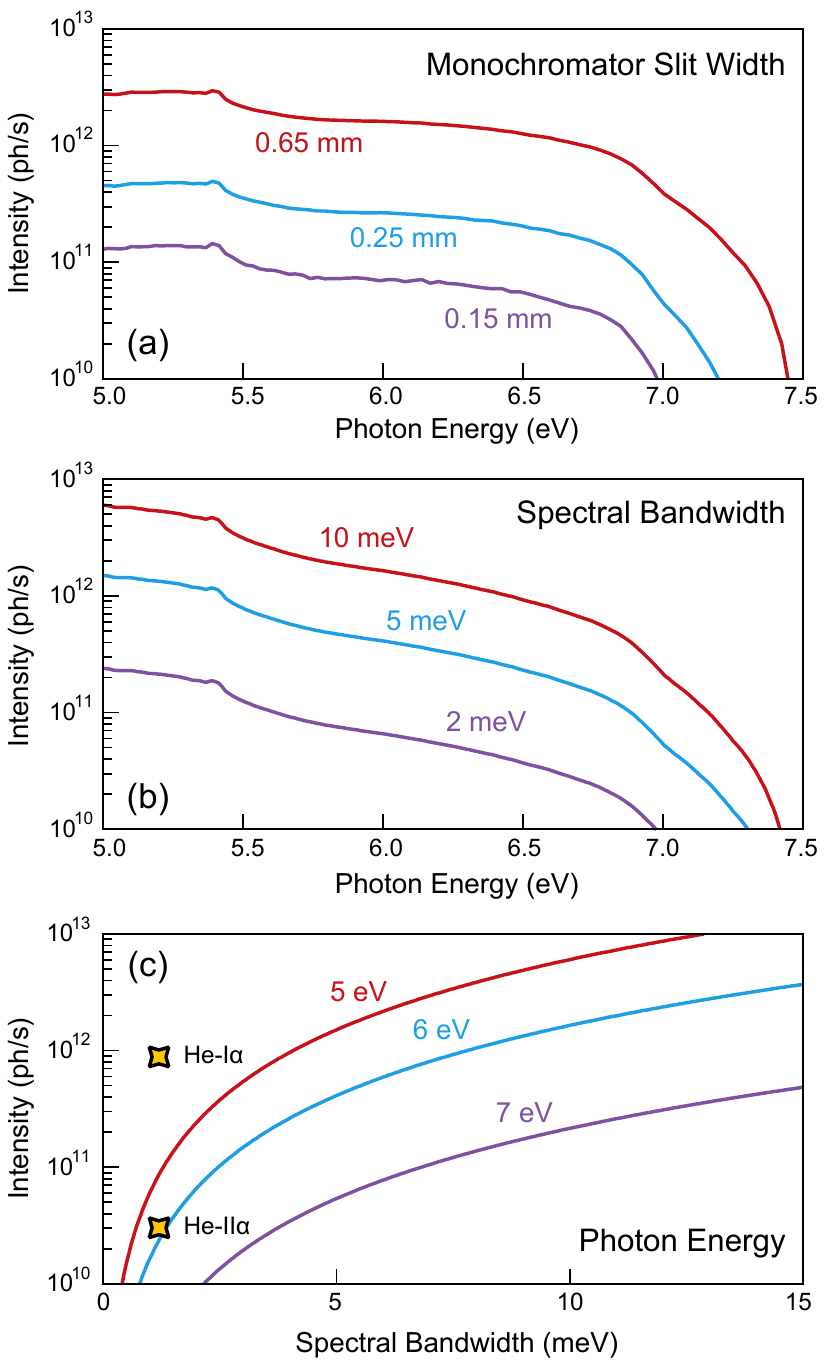}
\caption{\label{figureIntensity}Absolute intensity of the photon source.  (a) Experimentally measured intensity as a function of photon energy for a series of fixed monochromator slit widths.  (b) Calculated intensity as a function of photon energy for a series of desired spectral bandwidths.  (c) Calculated intensity as a function of spectral bandwidth for a series of photon energies and for a conventional helium plasma lamp.}
\end{figure}

The most important characteristic of a photon source for photoemission is photon flux (for a given spectral bandwidth). The instrument described here provides a very bright source, rivaling that of conventional laboratory-based plasma lamps. Figure \ref{figureIntensity} shows the absolute intensity of the source as a function of photon energy for a series of monochromator slit widths, measured at the output of the monochromator with a calibrated silicon photodiode (AXUV-20, International Radiation Detectors, Inc.).  Also shown is the expected intensity as a function of spectral bandwidth (assuming the slits are adjusted at each energy in order to keep the bandwidth constant). Superimposed on Fig.\ \ref{figureIntensity}(c) is data for a helium plasma lamp (VUV5000, VG Scienta, Inc.). Although the spectral bandwidth of the helium lamp is narrow ($\sim$1 meV), the photon energy is limited to two primary lines (He-I$\alpha$ and He-II$\alpha$).

\subsection{Spectral bandwidth}

\begin{figure}[ht]
\includegraphics{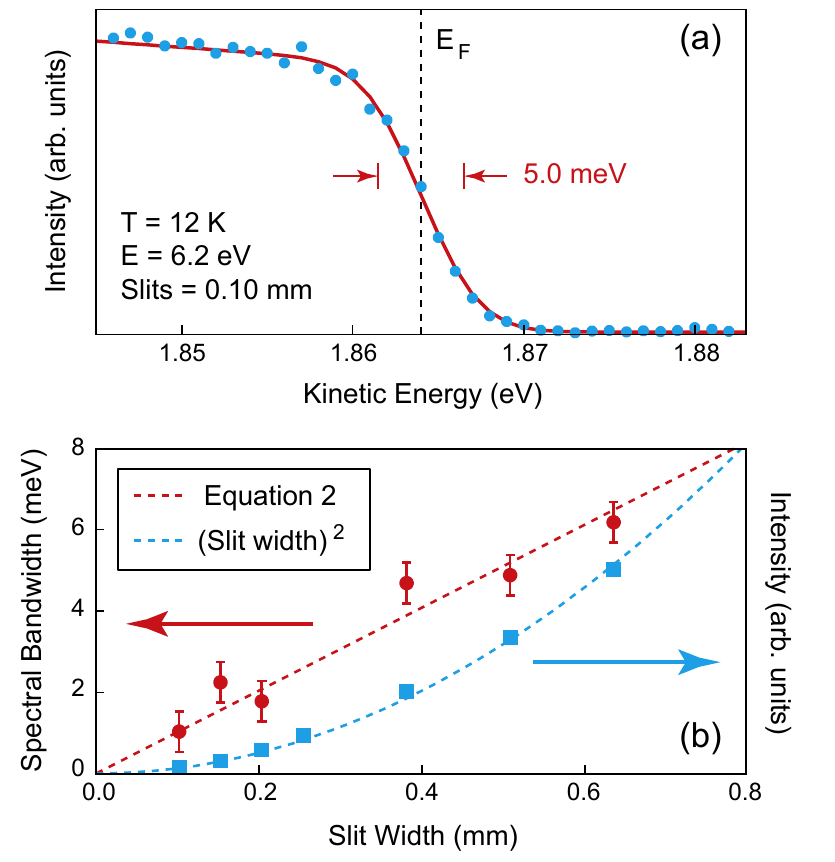}
\caption{\label{figureBandwidth}Spectral bandwidth and intensity of the photon source as a function of slit width.  (a) An angle-integrated energy distribution curve of polycrystalline gold taken at $E = 6.2$ eV and $T = 12$ K and with monochromator slits set to 0.10 mm. Also shown is a Fermi-Dirac distribution fit to the data, from which the spectral bandwidth can be extracted after deconvolving the broadening due to instrumental resolution (4.9 meV).  (b) Photoemission intensity and spectral bandwidth extracted using the procedure outlined in panel (a) as a function of monochromator slit width for $E = 6.2$ eV. The dashed red line shows Equation \ref{slitWidthEqn} and the dashed blue line shows the expected quadratic dependence of the intensity on slit width. Error bars in the spectral bandwidth come from uncertainties in the Fermi-Dirac distribution fit.}
\end{figure}

One feature of the photon source described here that is absent in conventional laboratory-based plasma lamps is the ability to adjust the spectral bandwidth of the source based on the requirements of the user: a significant intensity gain may be achieved by a moderate increase of the bandwidth of the light. This is especially useful, for example, when measuring momentum-space ARPES maps, as a large number of measurements must be taken, but each of which can be acquired with a somewhat lower energy resolution. Figure \ref{figureBandwidth} shows the measured spectral bandwidth and intensity of the photon source as a function of monochromator slit width. The bandwidth is determined by measuring the angle-integrated Fermi step of freshly evaporated polycrystalline gold and deconvolving the broadening due to instrumental resolution (4.9 meV) and finite temperature (12 K). The measured spectral bandwidth agrees well with Equation \ref{slitWidthEqn}, and the photoemission intensity follows the expected quadratic dependence on slit width due to the larger entrance slit and the wider bandwidth.

\subsection{Spot size}

\begin{figure}[ht]
\includegraphics{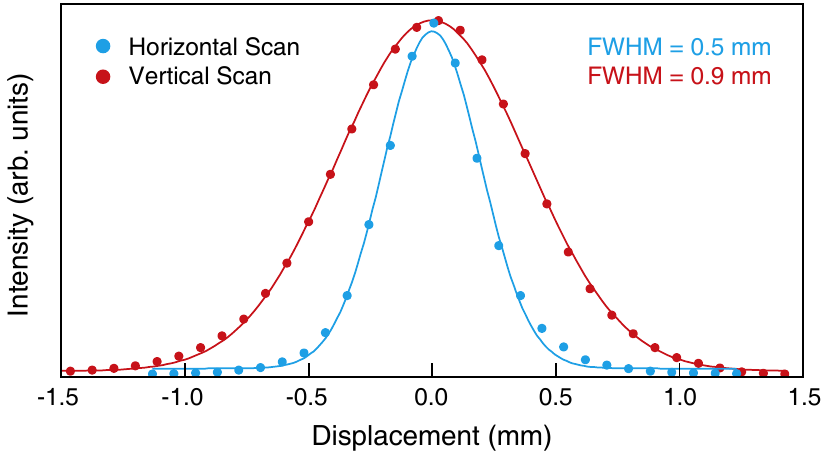}
\caption{\label{figureSpotSize}Size of the xenon plasma image formed at the monochromator entrance slit by the OAP assembly (3$\times$ magnification of the actual plasma). Solid lines are fits to a Gaussian function, from which a horizontal spot size of 0.5 mm (FWHM) and a vertical spot size of 0.9 mm (FWHM) are extracted.}
\end{figure}

Another important characteristic of a light source is a narrow spot size. This is required because samples studied by ARPES are usually small ($\sim$1 mm$^2$). If the spot size of the lamp is too large, most of the photons will contribute to a background signal, possibly overwhelming the signal from the sample itself. In addition, smaller spot sizes allow for higher momentum resolution in ARPES experiments. Figure \ref{figureSpotSize} shows the size of the xenon plasma image formed at the entrance slit of the monochromator by the OAP assembly. A full width at half maximum (FWHM) of 0.5 mm (horizontal direction) and 0.9 mm (vertical direction) are measured by the photodiode. Because the OAP assembly has a 3$\times$ magnification, we infer the xenon plasma to be roughly 0.17 mm by 0.3 mm within the bulb. For ARPES experiments, the slits of the monochromator usually stay within the range 0.1 -- 0.5 mm, which is on the order of the xenon lamp spot size at the entrance slit. As discussed above, a pair of lenses form an image of the exit slit at the sample position with a magnification of 2$\times$. The resulting spot size on the sample will thus be twice the slit width (0.2 -- 1.0 mm) in the horizontal direction and 1.8 mm in the vertical direction. A fixed vertical aperture, however, may be placed at the exit slit in order to reduce the vertical dimension.

\section{Photoemission}

\begin{figure}[ht]
\includegraphics{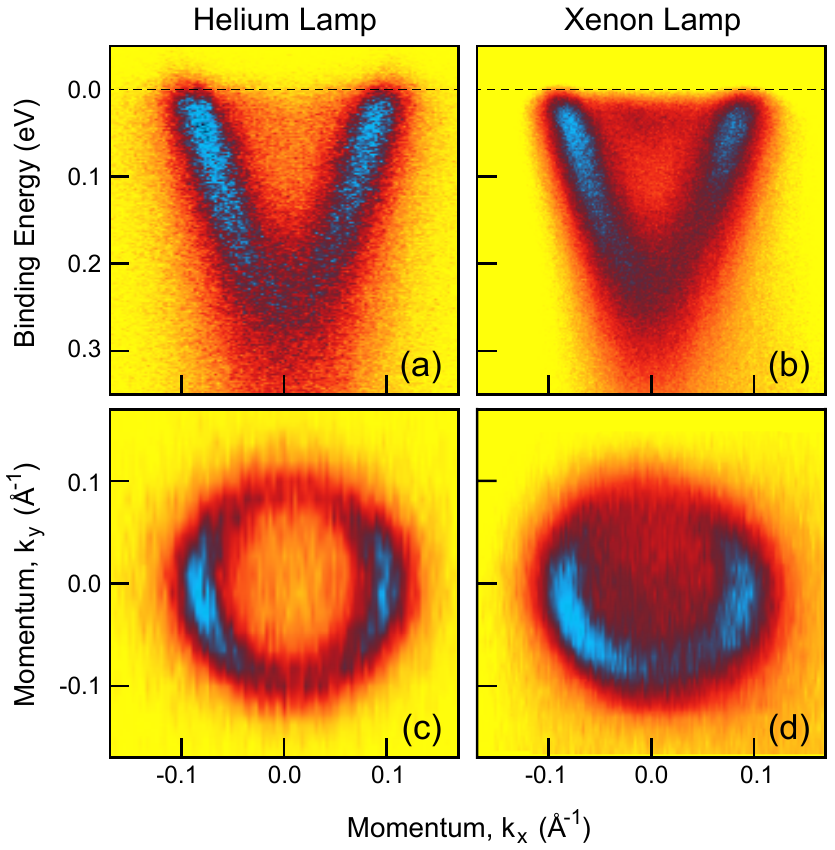}
\caption{\label{figureARPES}Comparative ARPES data showing the surface electronic structure of CdO using a conventional helium lamp (He-I$\alpha$, 21.2 eV) and the xenon photon source ($E = 6.2$ eV, 20 meV spectral bandwidth) at $T \approx 100$ K. The inner electron pocket is not visible in the helium map due to matrix element effects, but can be detected with the xenon light source.  (a),(b) Energy--momentum cuts through the center of the electron pocket ($k_y = 0$). For each photon source, data was accumulated for 60 minutes.  (c),(d) Fermi surface maps generated by integrating spectral weight within $\pm$30 meV of the Fermi energy. The He-I$\alpha$ map in panel (c) was acquired over a span of 160 minutes whereas the map in panel (d) taken with the xenon source required only 60 minutes.}
\end{figure}

To demonstrate the usefulness of the xenon photon source, we present ARPES data of a two-dimensional electron gas on the surface of CdO(001), which manifests itself as a pair of small concentric electron pockets centered at $k_x = k_y = 0$ in momentum space \cite{piper2008,king2010}. The CdO sample was grown by metalorganic vapor phase epitaxy on a sapphire substrate \cite{zunigaperez2004} and measures $\sim 10\times5$ mm$^2$. In order to produce a clean surface for photoemission, the sample was annealed in vacuum at 600 $^\circ$C for one hour before measurement. Figure \ref{figureARPES} compares energy--momentum cuts and Fermi surface maps of the surface electronic structure of CdO using a conventional helium plasma lamp (He-I$\alpha$) and using the xenon light source described in this paper. The Fermi surface maps were generated by integrating spectral intensity within $\pm$30 meV of the Fermi level; an energy resolution much smaller than this energy scale is therefore unnecessary, and the ability to increase the photon flux by widening the spectral bandwidth is desirable. This feature enabled the Fermi surface map to be acquired over a span of only 60 minutes with the xenon source, while 160 minutes was required to obtain a similar signal-to-noise ratio using the helium lamp. Both the helium and xenon plasmas are nominally unpolarized. In each case, however, reflection off the monochromator grating and optical elements partially polarizes the photons. This can account for the slight difference between the symmetry of the Fermi surface map intensity of Fig.\ \ref{figureARPES}(c) and Fig.\ \ref{figureARPES}(d). Photoemission matrix elements, which depend sensitively on photon energy, also play a role in the polarization dependence of the measured intensity.

\section{Concluding remarks}

Future improvements to the photon source discussed in this paper include the possibility of controlling the polarization of the emitted light A MgF$_2$ double Rochon prism \cite{steinmetz1967} or a triple-reflector Brewster polarizer \cite{hass1978} placed after the lens at the monochromator exit slit can linearly polarize the light before it enters the vacuum chamber. By rotating the prism or reflectors, one may control the direction of polarization of light incident upon the sample. A quarter waveplate operating in the ultraviolet wavelength range (for example, birefringent CaF$_2$ \cite{burnett2001}) can transform the linear polarization into circular polarization.

In conclusion, we have presented a novel photon source that is comprised of a laser-driven xenon plasma lamp coupled to a Czerny-Turner monochromator. We believe that this device will be a valuable addition to laboratory-based light sources for angle-resolved photoemission spectroscopy.

\begin{acknowledgments}
We acknowledge helpful discussions with H.\ Padmore and F.\ Baumberger. Photoemission was performed on a CdO sample provided by C.\ McConville and grown by V.\ Mu\~{n}oz-Sanjos\'{e}, and we thank D.\ G.\ Schlom for allowing us to utilize his vacuum chamber for sample preparation. This work was supported by the National Science Foundation through DMR-0847385 and the MRSEC program under DMR-1120296 (Cornell Center for Materials Research). E.J.M.\ acknowledges support from an NSERC PGS.
\end{acknowledgments}

\end{document}